\documentclass[twocolumn,showpacs,superscriptaddress,preprintnumbers,amsmath,amssymb,prc]{revtex4}

\usepackage{graphicx}
\usepackage{dcolumn}
\usepackage{bm}

\usepackage{color}
\usepackage{CJK}

\begin{document}
\begin{CJK*} {GB} {gbsn}

\title{Two-proton momentum correlation from photodisintegration of  $\alpha$-clustering light nuclei in the quasi-deuteron region}
\author{Bo-Song Huang (»Æ²ªËÉ)\footnote{Email: huangbosong@sinap.ac.cn}}
\affiliation{Key Laboratory of Nuclear Physics and Ion-Beam Application (MOE), Institute of Modern Physics, Fudan University, Shanghai 200433, China}
\affiliation{Shanghai Institute of Applied Physics, Chinese Academy of Sciences, Shanghai 201800, China}
\author{Yu-Gang Ma (ÂíÓà¸Õ)\footnote{Email: mayugang@fudan.edu.cn}}
\affiliation{Key Laboratory of Nuclear Physics and Ion-Beam Application (MOE), Institute of Modern Physics, Fudan University, Shanghai 200433, China}
\affiliation{Shanghai Institute of Applied Physics, Chinese Academy of Sciences, Shanghai 201800, China}

\begin{abstract}
The proton-proton momentum correlation function is constructed in three-body photo-disintegration channels from $^{12}$C and $^{16}$O targets in the quasi-deuteron regime
within the framework of an extended quantum molecular dynamics model. Using the formula of Lednicky and Lyuboshitz (LL) for the momentum correlation function, we obtain a proton-proton momentum correlation function for the specific three-body photon-disintegration channels of $^{12}$C and $^{16}$O targets, which are assumed to have different initial geometric structures, and extract
their respective emission source sizes for the proton-proton pair. The results demonstrate that constructing a proton-proton momentum correlation is feasible in photo-nuclear reactions, and it is sensitive to the initial nuclear structure. For future experimental studies investigating the $\alpha$-clustering structures of light nuclei, the present work can be used to shed light on the performance and correlation function analysis of ($\gamma$,pp) or (e,$e'$pp) reactions. 
\end{abstract}

\date{\today}


\maketitle

\section{Introduction}

The $\alpha$-clustering state plays a fundamental role in nuclear structure physics and nuclear astrophysics, as it is crucial for understanding both the process of nucleosynthesis and the abundance of elements~\cite{Ikeda,Greiner,Ortzen,Freer,JBN,Nature,THSR,An,An2}. For nuclei with $Z \le16$, the mean field effect is insufficiently strong to break cluster structures at low temperatures. Therefore, clustering behavior can be observed at excited states or even in the ground state.
For target nuclei such as $^{12}$C and $^{16}$O (as considered in this study), $\alpha$-clustering structures have been extensively discussed \cite{Ortzen}. $^{12}$C is of great interest because of its three-$\alpha$ clustering structure, which can be involved in astrophysical nucleosynthesis with its Hoyle state \cite{Hoyle}. $^{16}$O seems more ambiguous in its configurations. With a four-$\alpha$-clustering structure, a chain configuration was predicted by the Skyrme Cranked Hartree-Fock method~\cite{Ichikawa}, and a tetrahedral structure as a ground state was predicted based on the chiral nuclear effective field theory~\cite{Suhara}. Different geometrical shapes of the $\alpha$-clustering nuclei can induce rich properties of structure and reaction ~\cite{Schuck,16O_chain,CFT1,CFT2,D3,Zhou,CaoXG-2019,NST,Enyo}. Some probes have been presented as sensitive observables to geometrical shapes of clustering nuclei. For instance, giant dipole resonance (GDR) displays corresponding characteristic spectra for different $\alpha$-clustering configurations of $^{12}$C and $^{16}$O~\cite{W.B.He}. Collective observables show significant differences among various $\alpha$-clustering structures in heavy-ion collisions \cite{Guo,ZhangS,XuZW}. However, these probes are still limited and 
more probes are expected in the future. 
In this context, we suggest using the proton-proton ($p-p$) momentum correlation function to investigate different $\alpha$-clustering structures of $^{12}$C and $^{16}$O.

However, a photo-nuclear reaction is involved in the initial nuclear excitation process with incident high energy photons, which then induce phenomena such as nuclear resonance fluorescence, photo-disintegration, and photo-fission. This has been investigated for several decades and is considered a critical process for understanding the nuclear structures and fundamental dynamics of nucleonic systems. In particular, with the availability of high-quality monochromatic photon beams generated by the tagged photon technique or laser-electron Compton backscattering $\gamma$ sources ~\cite{Nuclear photonfissility,Duke,SIOM,ELI,SINAP, Amano}, using photon beams to investigate the behaviors of hadrons in a nuclear medium is very helpful. Different from the traditional ion beams, photon probes are elementary and non-hadronic and thus enable us to obtain information about the nuclear medium.
In the past decades, low energy-photon beams were mainly applied, for example, to studies on the giant dipole resonance (GDR) with 15--40 MeV photons \cite{photo_16O}. When the photon energy is higher than the GDR region and reaches approximately 140 MeV, the wavelength of the photons is typically smaller than the size of the nucleus, which is close to the size of the deuteron. To address this region, the quasi-deuteron (QD) absorption mechanism has been introduced \cite{J.S.Levinger0}. It is indicated that the photo-absorption of one proton-neutron ($p-n$) pair in the nucleus is dominant in this region, and therefore this process provides a tool for the study of nucleon-nucleon (NN) correlation in the nucleus. The $p-n$ correlation in $^{12}$C has been studied using the two-nucleon knockout reaction in the QD region ~\cite{np}. In a recent work, we investigated the photo-nuclear reactions of $^{12}$C and $^{16}$O with different $\alpha$-clustering structures in the QD region~\cite{huang12C,huang16O}, and found that some properties of ejected neutrons and protons are sensitive to the geometric structure of $\alpha$-clustering in a three-body decay channel. As a further step, we can imagine that a two-proton decay channel might be another useful probe for structures of $\alpha$-clustering nuclei and is thus a major task of the present work.

The momentum correlation function of two protons that are emitted through a final state interaction can be calculated by the Hanbury-Brown and Twiss (HBT) method, which is an intensity interferometry technology. The method was initially applied in the 1950s to stellar astronomy to measure the angular diameter of bright visual stars from coherent photon beams~\cite{Hanbury}. Later, this method was widely applied to elementary physics, such as in the 1960s for nuclear collisions at intermediate and high energy~\cite{Goldhaber}. It was demonstrated that two-particle correlations can be used as an estimation of the space-time dimensions of the emission region as well as a method to determine the form of short-range interaction potential.
Thus far, the nucleon-nucleon correlation function has been applied to investigate the heavy-ion collision dynamics at intermediate energy in the framework of different transport models~\cite{myg,cxg,wtt,wtt2}. In ultra-relativistic nuclear collisions, the first measurement of two-antiproton interaction was realized by analyzing the momentum correlation function between antiprotons, and the scattering length and effective range for the antiproton interactions were quantitatively extracted experimentally~\cite{STAR,ChenJH}. The same method was also proposed to search for new exotic hadron candidates (e.g., a possible dibaryon candidate $N\Omega$ \cite{Neha,WangFan,ShenPN} and a new antimatter nucleus $^4\overline{L}i$ \cite{Xi}.
Furthermore, this method has been applied to study some light nuclei with exotic structures, including proton-rich nuclei ($^{22}$Mg and $^{23}$Al, etc. \cite{Ma2015,Fang2016,Wang2018,Wang2018B}) and neutron-rich nuclei ($^{6}$He, $^{11}$Li, and $^{14}$Be, etc. \cite{Marques,Wei}).
Thus, to extend the HBT technique to light $\alpha$-conjugate nuclei and to examine the properties of the exotic structures are a natural consideration.

In this study, using a transport model (the extended quantum molecular dynamics (EQMD) model \cite{MARUYAMA}), we calculate a two-proton momentum correlation function for photo-disintegration at an incident photon energy of approximately 100 MeV. Using a QD mechanism, we demonstrate the feasibility of constructing the momentum correlation function for the emitted protons from a three-body photo-disintegrated channel using the LL formula. We then extract the emission source sizes for different $^{12}$C and $^{16}$O configurations.

The remainder of this paper is organized as follows. Methods of calculations are presented in Section 2, which includes three parts: a brief introduction to EQMD model, the process of QD absorption, and the LL analytical method. In Section 3, we present the main
results and discussion, which include the reliability check for our model, proton-proton momentum correlation functions, and the deduced source sizes for different $\alpha$-clustering structures of $^{12}$C and $^{16}$O with 100-MeV incident photons. In addition, an energy dependence of momentum correlation functions of $^{16}$O with the linear four-$\alpha$ structure and the corresponding source sizes are presented. The results demonstrate that the proton-proton momentum correlation function is sensitive to different $\alpha$-clustering structures of $^{12}$C and $^{16}$O. Therefore, this work can be used to shed light on future experimental studies in photo-nuclear facilities. Finally, a summary is provided in Section 4.

\section{Methods of calculations}

\subsection{EQMD model }

The quantum molecular dynamics (QMD)-type model~\cite{J.Aichelin,J.Aichelin2} have been extensively applied in dealing with fragment formation and correlation in heavy ion collisions at intermediate energy \cite{J.Aichelin2,C.Hartnack,C.Hartnack1,FengNST}.
However, descriptions of the ground state of the nucleus have not been sufficiently accurate for the QMD-type model, because the phase space obtained from Monte Carlo samples is typically not at the lowest point of energy. 
To solve this problem, an extended version of QMD (EQMD) has been developed ~\cite{MARUYAMA} and is used in our calculation.

Two features are introduced in the EQMD compared with the standard QMD. To cancel the zero-point energy caused by the wave packet broadening in the standard QMD, the cooling process can be used to maintain the mathematical ground state. However, the Pauli principle is then broken. Unlike in the standard QMD model, Fermi statistics are not satisfied in the EQMD because nucleons are not antisymmetrized. However, repulsion between identical nucleons is phenomenologically considered by a repulsive potential ~\cite{A.Ohnishi} known as a Pauli potential. As a result, saturation properties and $\alpha$-clustering structures can be obtained after energy cooling in the EQMD
model \cite{W.B.He}.
 Another feature is that the EQMD model treats the width of each wave packet as a dynamic variable~\cite{P.Valta}. The wave packet of the nucleon is taken in a Gaussian-like form as follows:
\begin{multline}
\phi_{i}(r_{i}) = \bigg(\frac{v_i+v^{*}_{i}}{2\pi}\bigg)^{3/4}exp\bigg[-\frac{v_{i}}{2}(\vec{r}_{i}-\vec{R}_{i})^{2} +\frac{i}{\hbar}\vec{P}_{i}\cdot \vec{r}_{i}\bigg],
\end{multline}
where $\vec{R}_{i}$ and $\vec{P}_{i}$ are the centers of position and momentum of the $i$-th wave packet, and the $v_{i}$ is the width of the wave packets, which can be presented as
${v_i} = {{1/{\lambda _i}}} + i{\delta _i}$, where $\lambda_i$ and $\delta_i$ are dynamic variables.
The ${v_i}$ of the Gaussian wave packet for each nucleon is dynamic and independent.

The Hamiltonian of the entire system is written as follows:
\begin{multline}
H = \left\langle \Psi \mid \sum_{i} -\frac{h^{2}}{2m}\bigtriangledown^{2}_{i}-\widehat{T}_{c.m.}+\widehat{H}_{int} \mid\Psi \right\rangle\\
\\=\sum_{i}\bigg[\frac{{\vec{P}_i}^2}{2m}+\frac{3\hbar^{2}(1+\lambda^{2}_{i}\delta^{2}_{i} )}{4m\lambda_{i}} \bigg]-T_{c.m.}+H_{int},
\label{eq_H}
\end{multline}
where $T_{c.m.}$ is the zero-point center-of-mass kinetic energy ~\cite{A.Ono} and
$H_{int}$ is the interaction potential in the form of
\begin{equation}
H_{int} = H_{Skyrme} + H_{Coulomb} + H_{Symmetry} + H_{Pauli},
\end{equation}
where the Pauli potential $H_{Pauli} = \frac{c_{ P}}{2}\sum_{j}(f_{i}-f_{0})^{\mu}\theta(f_{i}-f_{0})$
with $f_{i}$ is defined as an overlap of the $i$-th nucleon with other nucleons that have the same spin and isospin.

In the present work, we simulate the photo-absorption and photo-disintegration in the EQMD model with the obtained configurations for $^{12}$C and $^{16}$O and treat three-body decay properties.


 \subsection{Process of QD absorption}

 Photo-nuclear reaction has been used as a probe for nuclear structures in describing sensitive observations within the EQMD model. In this section, we clarify the photo-absorption process by a QD mechanism. For details, a single proton-neutron pair in a single $\alpha$ cluster of given $\alpha$-conjugate nuclei is bombarded with incident photons at energy in the QD region, and then the nucleus is excited by the absorption process and enters the transport process to the final state, and finally leads to particle ejection. In this study, a three-body decay channel with two protons and one residual nucleus was our only focus (where other decay channels are not discussed). The phase space information of the emitting protons is taken as the input for our correlation function calculations using the LL method, which is briefly introduced later.

A proton-neutron pair inside the nucleus can be treated as a QD when incident photons are in intermediate energy of approximately 70--140 MeV. In this case, the photon absorption mechanism plays a dominant role, and the QD photo-disintegration reaction is considered based on Levinger's QD model ~\cite{J.S.Levinger}, where the latter employs an impulse approximation method that considers the remaining nucleons and the correlated proton-neutron pair act as spectators after incident photons have been absorbed.

In the calculation, different configurations of $^{12}$C and $^{16}$O obtained from the cooling process with the Pauli potential in the EQMD model are considered as the inputs of the phase space. For nuclei composed of N-$\alpha$ clusters, we can simplify our consideration through an absorption process ($\gamma$,$^{4}$He) in which an $\alpha$ cluster inside the target is chosen randomly. We then assume the remaining two nucleons and absorbed QD inside this and other clusters in the nucleus are spectators. This is because the spatial separation between $\alpha$ clusters is much greater than the distance between a pair of QDs in the EQMD frame. However, in the assumptions of other models such as microscopic cluster models, it becomes more complicated.
The kinetic process in our calculation is such that photon energy transfers to the proton-neutron pair of the chosen $\alpha$ cluster and its kinetic process is replaced by $^{2}$H($\gamma$, np). Whether the process occurs depends on the cross section of $^{2}$H($\gamma$,np) in each event by Monte Carlo sampling.
The cross section that uses this calculation is integrated from the angular-dependent formula of the proton of this reaction as fitted by Rossi {\it et al.} ~\cite{P.Rossi}, where the incident photon energy ranges from 20 to 440 MeV in the center-of-mass (CM) frame. More details can be found in the literature ~\cite{P.Rossi}.

Because only one $\alpha$ cluster interacts with photons in each photo-nuclear reaction event,
we select one proton-neutron pair inside an $\alpha$ cluster by Monte Carlo sampling according to the cross section formula of $^{2}$H($\gamma$,np). The total four-momentum in the system for the photon-absorption in the laboratory frame can be written as 
$ \vec{P}^{Lab}_{tot} = \vec{P}^{Lab}_{\gamma} + \vec{P}^{Lab}_{QD}$.
 We then translate the CM frame using the Lorentz boost. The total momentum of the system before absorption is 
$ \vec{P}^{cm}_{tot} = L(\beta) \vec{P}^{Lab}_{tot},$
where
$ \beta = {P}^{LAB}_{tot}/P^{LAB}_{tot}(0)$, $L(\beta)$ is the operation of the Lorentz transformation, and $\vec{P}^{Lab}_{tot}(0)$ is the total energy of the two-body system in the CM frame.

In terms of conservation of momentum and energy, the four-momentum of the outing proton-neutron pair of $^{4}$He($\gamma$, pn)d is written as 
 $E^{cm}_{p}  = E^{cm}_{n} = P^{cm}_{tot}(0)/2$ and
$\vec{P}^{cm}_{p}  = -\vec{P}^{cm}_{n} = \sqrt{m^{2}+(\vec{P}^{cm}_{tot}(0)/2)^{2}},$
 where the $m$ is the mass of the nucleon. The angular distribution of outgoing nucleons is obtained by the differential cross section of $(\gamma,np)$ using a Monte Carlo sampling of the $^{2}$H($\gamma$,p)n differential cross section. We assume that the incoming photons are randomly distributed in the $xy$ plane. We then choose this event when the incoming photon is inside the region of the QD total cross section.
After the initial process of $(\gamma,np)$ has been completed, the nucleus is excited, and the nucleon can be emitted through final state interaction (FSI).

\subsection{LL analytical method}

Through final state interaction, we can use the phase space information at the emission time to construct a momentum correlation function. Before demonstrating our results, we describe the HBT calculation using the LL method~\cite{Lednicky}.
The LL method is based on the principle that the correlation functions of identical particles when emitted at small relative momenta are determined by the effects of quantum-statistical symmetry of particles and the final-state interaction 
\cite{Koonin1977}. The correlation function can then be expressed through a square of the symmetrized Bethe-Salpeter amplitude averaged over the four coordinates of the emission particles and the total spin of the two-particle system, which represents the continuous spectrum of the two-particle state.
In this model, the FSI of particle pairs is assumed to be independent in the production process. Based on the conditions described in Ref. \cite{Lednicky1}, the correlation function of two particles can be written as
\begin{equation}
\textbf{C}\left(\textbf{k}^*\right) = \frac{\int
\textbf{S}\left(\textbf{r}^*,\textbf{k}^*\right)
\left|\Psi_{\textbf{k}^*}\left(\textbf{r}^*\right)\right|^{2}d^{4}\textbf{r}^*}
{\int
\textbf{S}\left(\textbf{r}^*,\textbf{k}^*\right)d^{4}\textbf{r}^*},
\end{equation}
where $\textbf{r}^* = \textbf{x}_{1}-\textbf{x}_{2}$ is the relative distance between the two particles at their kinetic freeze-out, $\textbf{k}^*$ is half of the relative momentum between two particles, $\textbf{S}\left(\textbf{r}^*,\textbf{k}^*\right)$ is the probability to emit a particle pair with given $\textbf{r}^*$ and $\textbf{k}^*$ ($i.e.$, the source emission function), and $\Psi_{\textbf{k}^*}\left(\textbf{r}^*\right)$ is the Bethe-Salpeter amplitude, which can be approximated by the outer solution of the scattering problem \cite{STAR}.
With the aforementioned limit, the asymptotic solution of the wave function of the two charged particles takes the following approximate expression:
\begin{multline}
\Psi_{\textbf{k}^*}\left(\textbf{r}^*\right) = e^{i\delta_{c}}\sqrt{A_{c}\left(\lambda \right)} \times\\
\left[e^{-i\textbf{k}^*\textbf{r}^*}F\left(-i\lambda,1,i\xi\right)+f_c\left(k^*\right)\frac{\tilde{G}\left(\rho,\lambda \right)}{r^*}\right],
\end{multline}
where $\delta_{c} =
$arg$\Gamma\left(1+i\lambda\right)$ is the Coulomb s-wave phase shift with $\lambda = \left(k^*a_c\right)^{-1}$ in which $a_{c}$ is the two-particle Bohr radius, $A_c\left(\lambda \right) = 2\pi\lambda \left[\exp\left(2\pi\lambda \right)-1\right]^{-1}$ is the Coulomb penetration factor, and its positive (negative) value corresponds to the repulsion (attraction). In addition, $\tilde{G}\left(\rho,\lambda \right) = \sqrt{A_{c}\left(\lambda \right)}\left[G_0\left(\rho,\lambda \right)+iF_0\left(\rho,\lambda \right)\right]$ is a combination of regular $\left(F_0\right)$ and singular $\left(G_0\right)$ s-wave Coulomb functions \cite{Lednicky2009,Lednicky2008}, and $F\left(-i\lambda,1,i\xi\right) = 1+\left(-i\lambda\right)\left(i\xi\right)/1!^{2}+\left(-i\lambda\right)\left(-i\lambda+1\right)\left(i\xi\right)^{2}/2!^{2}+\cdots$ is the confluent hypergeometric function with $\xi = \textbf{k}^*\textbf{r}^*+\rho$, $\rho=k^*r^*$.

The s-wave scattering amplitude ($f_c\left(k^*\right)$) is renormalized by the long-range Coulomb interaction. It is expressed as follows:
\begin{equation}
f_c\left(k^*\right) = \left[ K_{c}\left(k^*\right)-\frac{2}{a_c}h\left(\lambda \right)-ik^*A_{c}\left(\lambda \right)\right]^{-1},
\end{equation}
where $h\left(\lambda \right) = \lambda^{2}\sum_{n=1}^{\infty}\left[n\left(n^2+\lambda^2\right)\right]^{-1}-C-\ln\left[\lambda \right]$ with the Euler constant $C$ = 0.5772.
$K_{c}\left(k^*\right) = \frac{1}{f_0} + \frac{1}{2}d_0k^{*^2} + Pk^{*^4} + \cdots$ is the effective range function in which $d_{0}$ is the effective radius of the strong interaction, $f_{0}$ is the scattering length, and $P$ is the shape parameter. The parameters of the effective range function are important parameters that characterize the essential properties of the FSI and can be extracted from the correlation function measured experimentally \cite{Erazmus1994,Arvieux1974,STAR}.

\section{Results and discussion}
\begin{figure*}
\center
\includegraphics[scale=0.45]{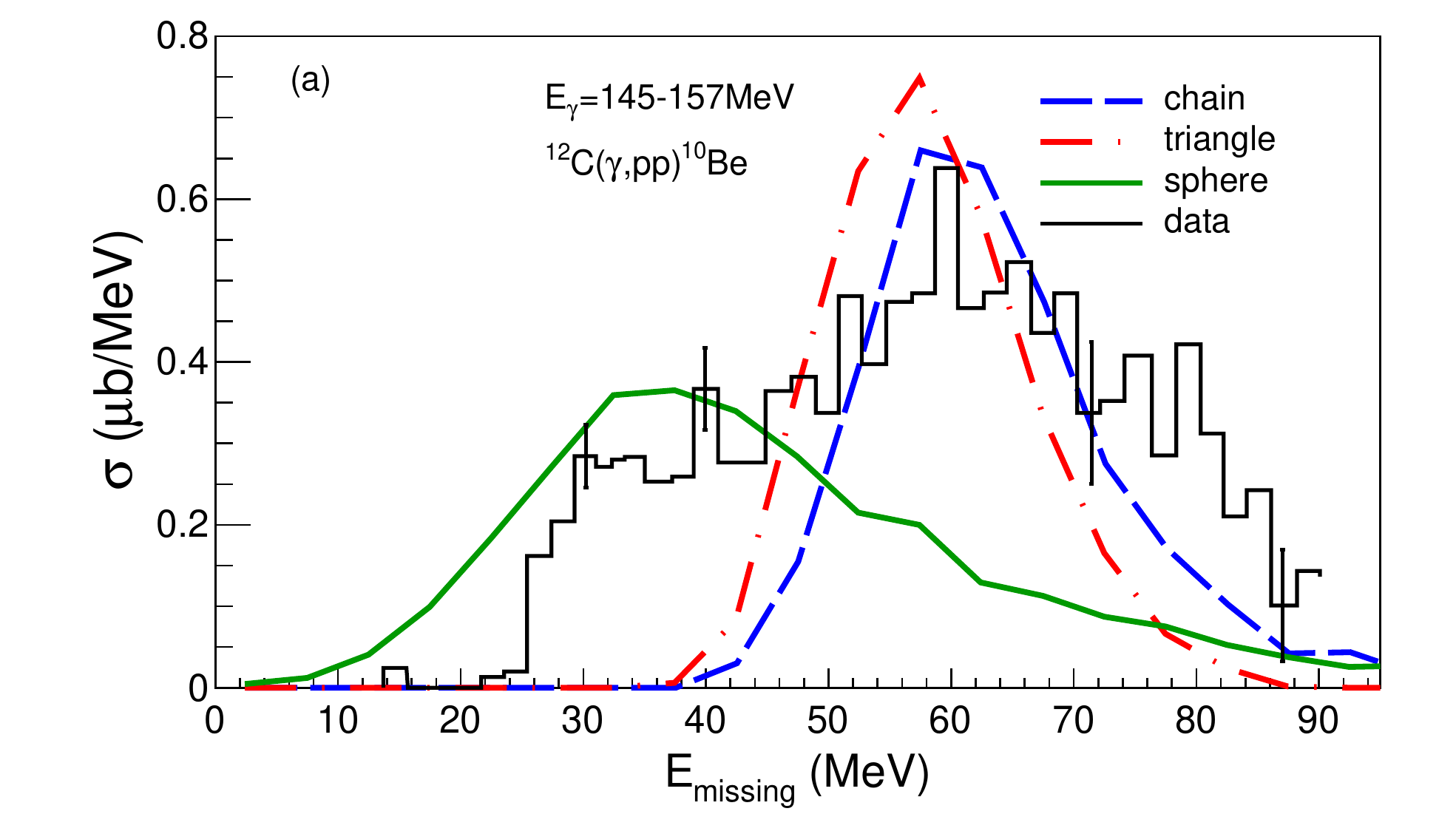}\includegraphics[scale=0.45]{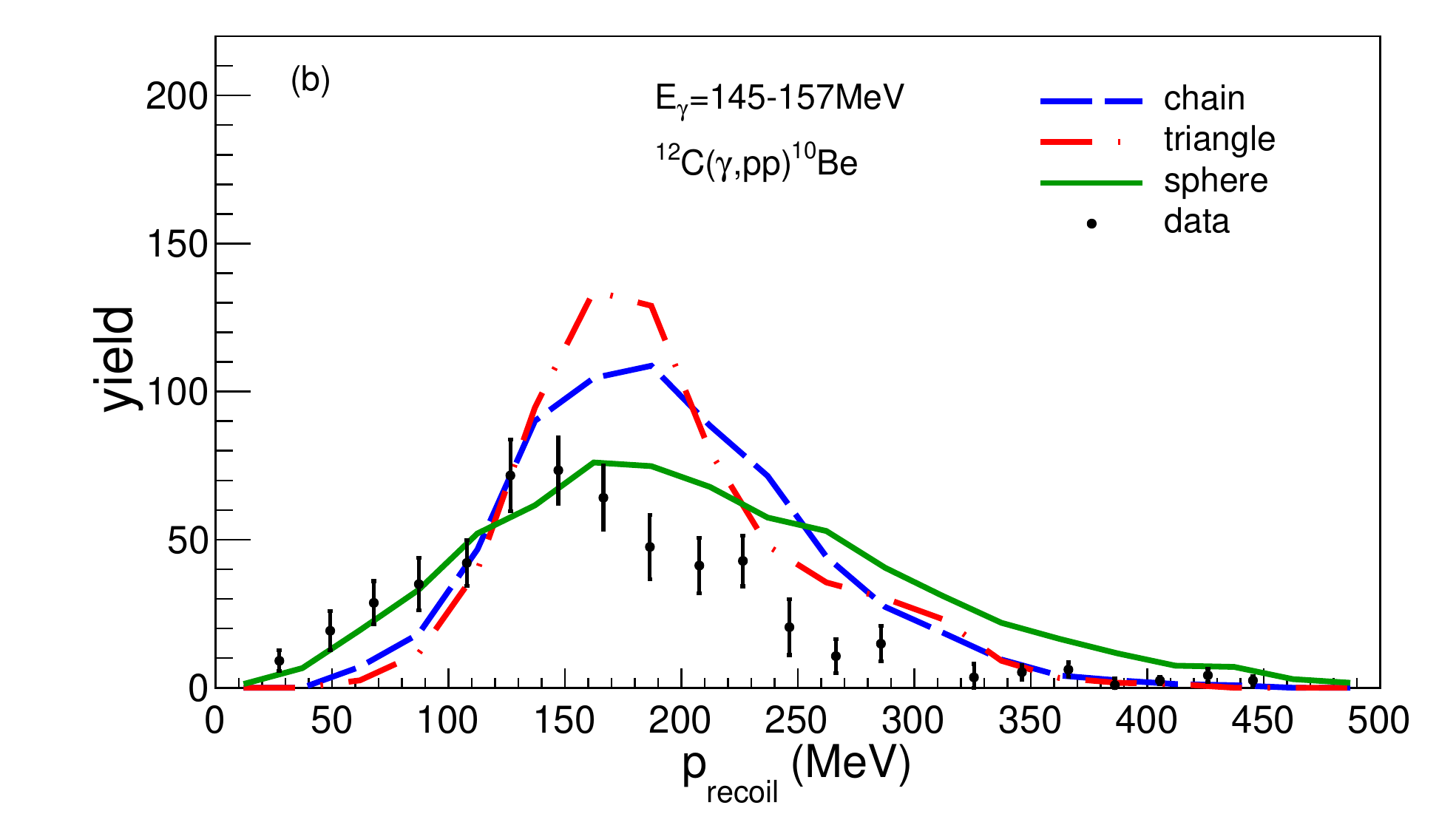}
\caption{Missing energy (a) and recoil momentum (b) spectra for $^{12}$C($\gamma$,pp) at 
E$_\gamma$ = 145--157 MeV. Note that the cut of E$_{miss}< $40 MeV is applied as the data \cite{MacNew} for $P_{recoil}$  (b).
Different lines represent different initial geometric configurations of $^{12}$C, as indicated in the insert. Please see the corresponding literature for details.}
\label{fig_comp}
\end{figure*}

We quantitatively compared the data to verify the model's reliability (e.g., by checking the recoil momentum spectrum and the missing energy spectrum). The recoil momentum is defined as $\vec{P}_{recoil} =\vec{P}_{\gamma} - \vec{P}_{p1} - \vec{P}_{p2}$, whereas the missing energy takes the form $E_{missing} = E_\gamma - T_{p1} - T_{p2} - T_{recoil}$. Here, $p1$ and $p2$ denote two emitted protons; $\vec{P}_{\gamma}$ is the momentum of the incident photon; $\vec{P}_{p1}$ and $\vec{P}_{p2}$ are the momenta of the two emitted protons; and $T_{p1}$, $T_{p2}$, and $T_{recoil}$ are the kinetic energies of the two protons and recoiled residue, respectively. In our previous work \cite{huang12C,huang16O}, we quantitatively compared the data for photo-$^{12}$C and -$^{16}$O reactions at E$_\gamma$ = 80--130 MeV \cite{McGeorge,McGeorge2}. The results demonstrated that $E_{missing}$ and $P_{recoil}$ spectra agreed well with the data. Here, we added a new example with the data comparison (i.e., the missing energy and recoil momentum spectra of $^{12}$C($\gamma$,pp)$^{10}$Be at E$_\gamma$ = 145--157 MeV). 
 Figure \ref{fig_comp} displays a comparison of our calculations with the data \cite{MacNew}
for $E_{missing}$ (a) as well as $P_{recoil}$ of $^{12}$C($\gamma$,pp)$^{10}$Be under the cut of $E_{missing}<40$ MeV (b). 
Note that the calculations shown in the figure were normalized with the same scale for comparison with the data. An observation of the $E_{missing}$ spectrum suggested that the addition of the sphere + triangle (or chain) could reproduce the spectra, thereby indicating a multi-configuration feature of the $^{12}$C nucleus. A review of the $P_{recoil}$ spectrum suggested that all three could give a broad peak position of $E_{missing}$ at approximately 150 MeV/c. However, for the width, the mixture of the sphere plus triangle may work well. We did not expect perfect fits for the data from our dynamic model, but the overall good agreement indicated that our model is capable of exploring more physics.

Based on model reliability, we investigated other observables such as the proton-proton momentum correlation function for photo-disintegrations of $^{12}$C and $^{16}$O with different $\alpha$-clustering configurations, which are obtained by cooling using the EQMD model \cite{W.B.He}. For a comparison with non-structured $^{12}$C and $^{16}$O, we used the Woods-Saxon nucleon distribution for both nuclei, which are tagged as spheres in the texts.  Many different photo-disintegration channels were derived from our full calculations (e.g., $^{12}$C($\gamma$,np)$^{10}$B and $^{16}$O($\gamma$,np)$^{14}$N are respective dominant channels
 in the process of $\gamma$+$^{12}$C and $\gamma$+$^{16}$O, which have roughly a 90$\%$ branching ratio for photo-disintegration from $\alpha$-clustering nuclei or roughly a 50--60$\%$ branching ratio from the Woods-Saxon spheric nucleus). However, in this study, we focused only on the three-body decay channel in the final state, which includes a residue and two protons, considering that effectively detecting neutrons in most experiments is difficult. In fact, the branching ratios for two-proton channel from each configuration are rare. Specifically, we found that for the $^{16}$O case, they were only
0.40$\%$, 0.70$\%$, 0.85$\%$, 1.30$\%$, and 5.13$\%$ for the
chain, kite, square, tetrahedron, and sphere configurations, respectively, with the total number of simulation events being 0.5 million. For the $^{12}$C case, they were only 0.45$\%$, 0.75$\%$, and 5.05$\%$ for the chain, triangle, and sphere configurations, respectively, with the total number of simulation events being 0.25 million. In these data, the chain configuration had the smallest two-proton emission branching ratio, whereas the spheric configuration had the largest. Later, we found that the increasing trend of the two-proton emission branching ratio from the chain, kite, square, tetrahedron, and sphere configurations was in line with the decreasing trend of the proton-proton emission source size or initial nuclear size. For details, please see Tables I and II.

Although the branching ratios were very small, the proton-proton momentum correlation functions could be reconstructed based on the phase space information. In fact, in our previous heavy-ion experiment, the two-proton emission probability was also very low. However, the proton-proton correlation can still be investigated \cite{Ma2015,Fang2016}. In this photo-nuclear reaction simulation, the final-state phase spaces of emitted protons were recorded after photo-absorption within the EQMD frame, which were taken as the inputs of the LL model. Before the correlation functions were calculated, we needed to know the emission times of two protons in the three-body exit channel. The times for nucleon emission were calculated starting from the beginning of photon absorption. When a proton-neutron (QD) pair inside the nucleus absorbed photon energy, it obtained higher kinetic energy and interacted with other nucleons. Through a method of nucleonic coalescence at each time step, the process can be roughly taken whereby the target ejects two protons and reorganizes other nucleons into a residue nucleus. We could track two emitted protons and obtain their emission times and then use the current emission time and phase space information as inputs for calculating the correlation functions in the LL model.

\subsection{Proton-proton momentum correlation functions for different $\alpha$-clustering structures}

\begin{figure*}
\center
\includegraphics[scale=0.45]{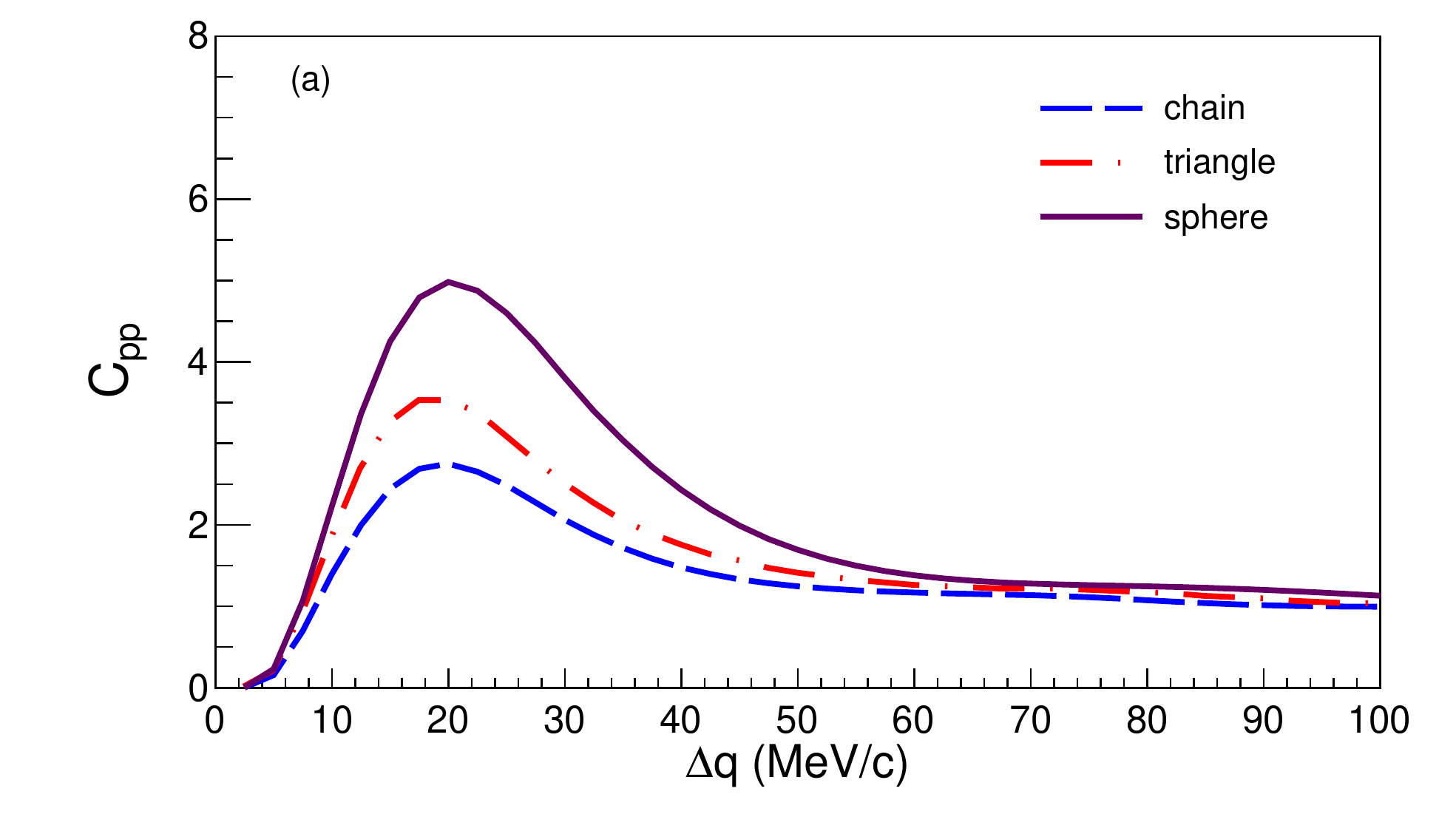}\includegraphics[scale=0.45]{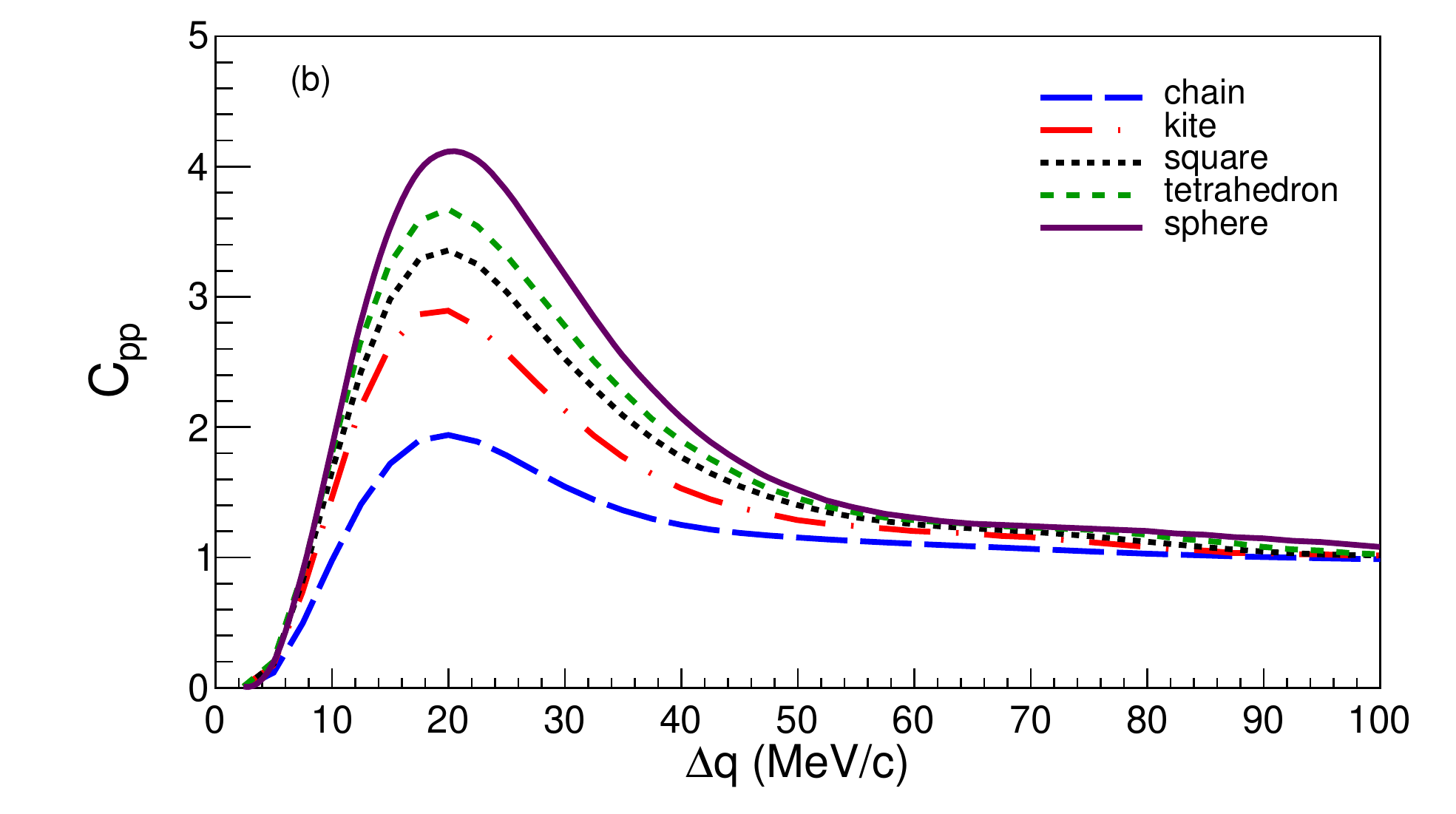}
\caption{Momentum correlation functions of two emitted protons from different initial $\alpha$-clustering structures of $^{12}$C (a) and $^{16}$O (b) bombarded with 100 MeV photons. Different lines represent different initial geometric configurations.}
\label{fig_Cpp_100MeV}
\end{figure*}

The calculations of the $p-p$ momentum correlation function for $^{12}$C($\gamma$,pp)$^{10}$Be and $^{16}$O($\gamma$,pp)$^{14}$C are presented in Fig. ~\ref{fig_Cpp_100MeV}. We can clearly see that the correlation functions show a dip at a smaller relative momentum ($\Delta q = |\vec{p_1}-\vec{p_2}|/2)$), which derived from the Coulomb repulsion, and a broad peak at approximately 20 MeV/c, which originated from the singlet proton-proton attractive interaction. It then tends to the unit at a larger $\Delta q$ because of the vanishing correlation. It is interesting that the correlation strength of $C_{pp}$ at approximately 20 MeV/c in Fig.~\ref{fig_Cpp_100MeV} is sensitive to the configuration structure, which indicates a different source size and/or emission time. For the $^{12}$C case, the spheric structure (i.e., the random nucleon distribution inside the nucleus) yields the largest $C_{pp}$, whereas the chain $\alpha$-clustering structure has the lowest correlation strength and the triangle $\alpha$-clustering is in between.
For the $^{16}$O case, the situations are similar but with a greater number of configurations (i.e., the spheric case displays the strongest correlation, the tetrahedron $\alpha$-clustering structure displays the second, the square and kite are in between, and the chain $\alpha$-clustering structure shows the weakest strength).

Fig.~\ref{fig_Cpp_100MeV} can be explained by the effective emission source size of the proton-proton from different nucleon distribution structures.
In the traditional interpretation of the $p-p$ HBT correlation, a correspondence exists between a strong correlation function and compact source size.
Because the chain structure has the largest size, its correlation function is the weakest, and it corresponds to the largest emission source size.
For the spheric nucleon distributions of $^{12}$C and $^{16}$O, the energy cooling process in the initialization of EQMD makes the nucleus very compact. It then shows the strongest HBT correlation strength. However,
the triangle structure of $^{12}$C and tetrahedron structure of $^{16}$O have very good symmetric structures. Accordingly, the correlation functions are the strongest, illustrating the most compact emission source size among all $\alpha$-clustering configurations.
The square and kite $\alpha$-clustering structures appear as the middle HBT peaks. The stronger peak in the case of the square indicates a smaller source size than in the kite case. Of course, this represents only a qualitative examination. Later, we extract the source size for different cases to support our judgments.

For the QD absorption mechanism in our calculation, a certain neutron-proton pair undergoes photon absorption, and the vast majority of initial neutrons and protons in the same cluster are finally emitted.
However, two-proton emission can still be observable despite its very low emission probability. The emission mechanism of two protons is as follows: the first proton is knocked out due to photo-absorption, and the second is primarily emitted through a knocked-out neutron exchanging with another proton in another $\alpha$ cluster. In this case, the time difference between two outgoing protons is much longer, which results in a decrease in correlation strength.

\begin{figure*}
\center
\includegraphics[scale=0.45]{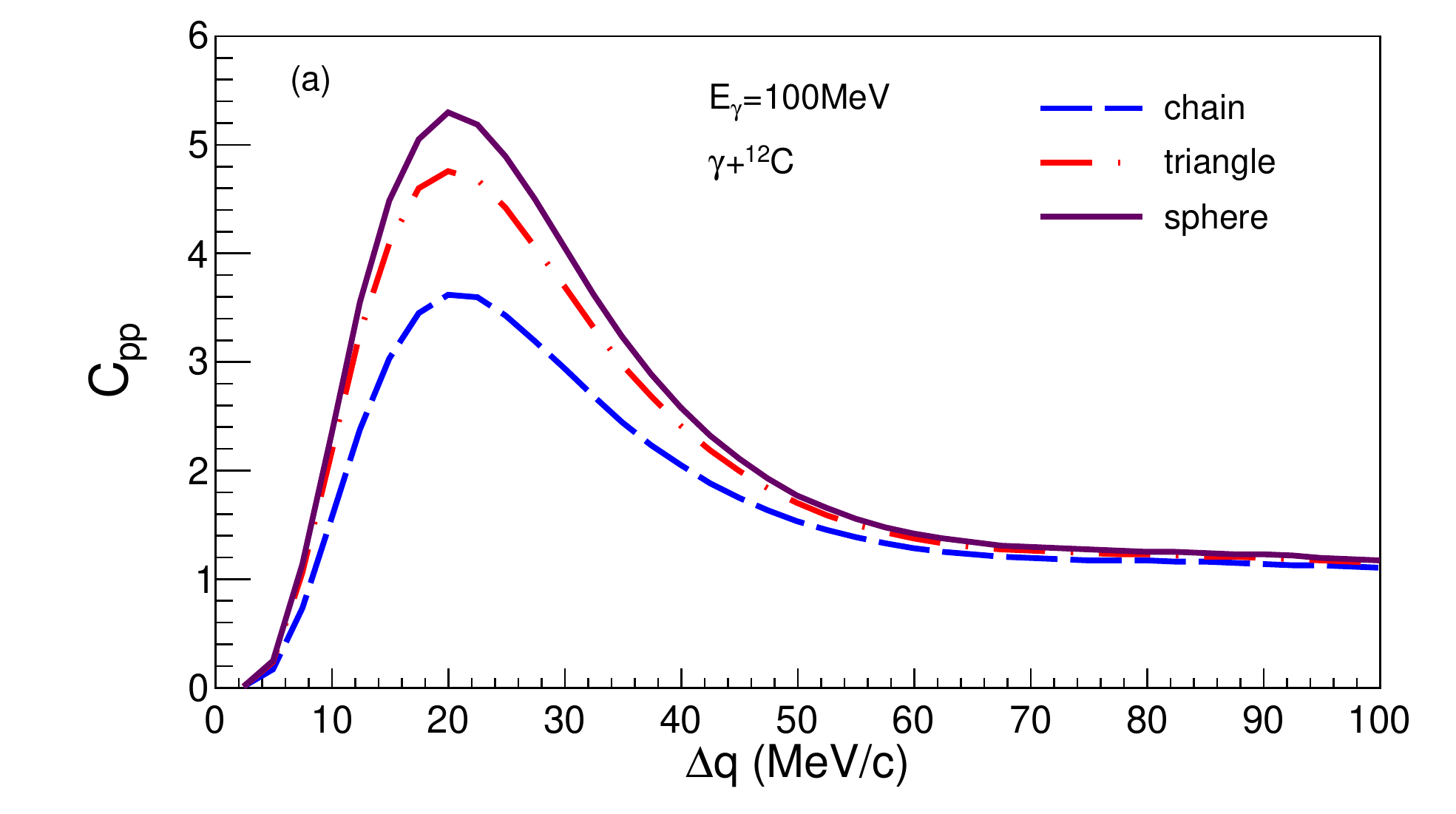}\includegraphics[scale=0.45]{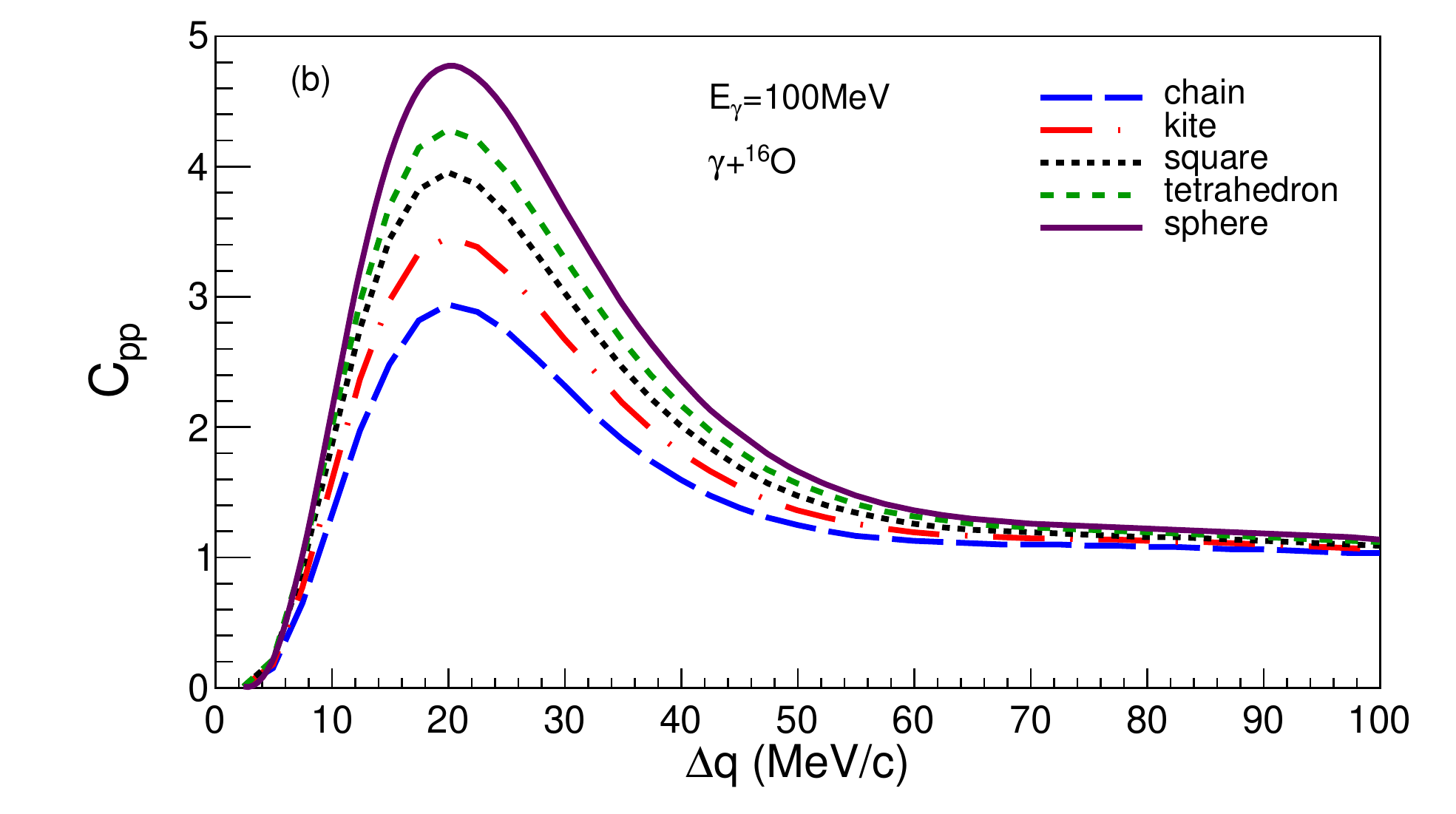}
\caption{Same as Fig.~\ref{fig_Cpp_100MeV} but with all emitted protons from all decay channels. }
\label{fig_allpp}
\end{figure*}

To verify such an ideal, we compare the proton-proton momentum correlation functions in which protons derive from all exit channels rather than only a two-proton-plus--residue channel. Fig.~\ref{fig_allpp} shows these results. In comparison with the correlation functions constructed from the two-proton emission channel (Fig.~\ref{fig_Cpp_100MeV}), the order of peak strength for different configurations does not change. However, the magnitudes for each configuration increase. The former illustrates that the proton-proton correlation method is actually a sensitive probe for different configurations through photo-nuclear reactions, regardless of two-proton emission channel or all proton channels. The latter illustrates that the effective emission source from those emitted protons are smaller and/or the time difference between two outgoing protons is relative shorter compared with the two-proton-emission case. This was explained in the previous paragraph.

\begin{figure}
\center
\includegraphics[scale=0.45]{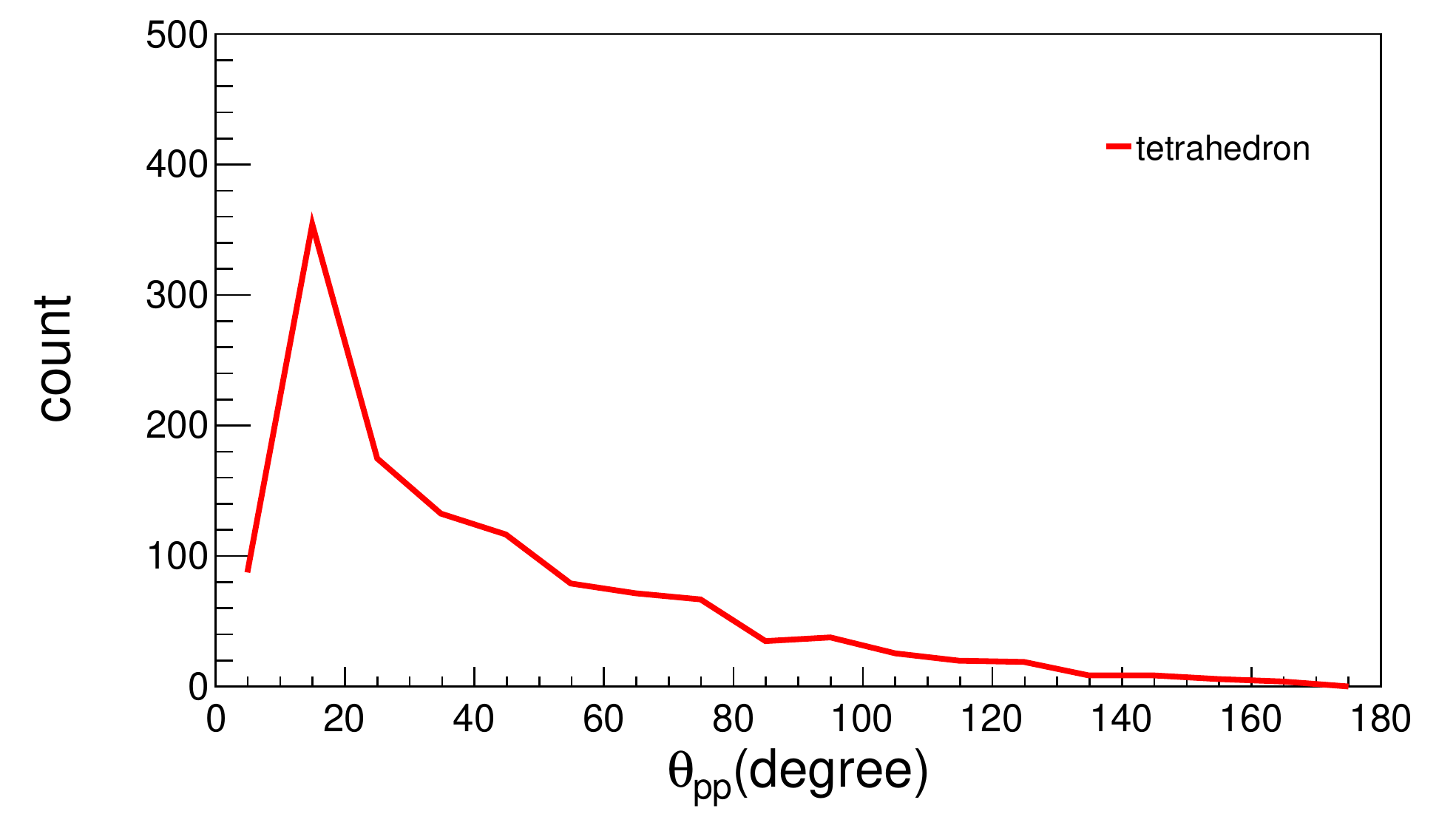}
\caption{Opening angular distribution between two emitted protons with a relative momentum cut between 15 and 25 MeV/c for the tetrahedron configuration of $^{16}$O.}
\label{fig_recol_p_12C}
\end{figure}

Although a strong correlation exists at approximately $\Delta q \sim $ 20 MeV/c in the momentum space, this correlation may emerge between the emission angle of protons.
To check this, we plotted Fig.~\ref{fig_recol_p_12C} for distribution of the opening angle between two emitted protons for the tetrahedron configuration of $^{16}$O, where a cut of $\Delta q$ is taken between 15 and 25 MeV/c. As expected, an evident peak emerges in the small angle range of approximately 20 degrees. This indicates a stronger smaller angle emission between the correlated two protons at $\Delta q \sim 20$ MeV/c \cite{Ma2015}, which is significantly different from the random emission scenario between two uncorrelated protons.

\subsection{Source sizes}

Before we discuss quantitative extraction of the emission source size from the proton-proton correlation function, it is helpful to determine the RMS radii for different initial nuclei, including $\alpha$-clustering configurations. Tables I and II show these results. It is obvious that the chain structure, which is extremely deformed, has the longest root mean square (RMS) radius, whereas the triangle or tetrahedron structure
 is more compact and spatially symmetric. In principal, the source size reflects proton occupancy of the space. Therefore, the chain structures of $^{12}$C and $^{16}$O demonstrate larger sizes, whereas the other configurations with more compact geometric space show smaller sizes. Table II shows that the difference between the RMS radii of the chain and kite configurations is more significant than the differences between other configurations. Thus, $C_{pp}$ is more distinguishable for its chain structure than in other configurations. By contrast, the square structure of $^{16}$O approximates the tetrahedron configuration in terms of spatial symmetry. These were also similar in terms of momentum correlation functions, as shown in Fig. ~\ref{fig_Cpp_100MeV}.

\begin{table}[]
\centering
\caption{RMS radius ($r_{RMS}$), binding energy ($E_{bind}/A$), HBT radius ($R_{source}$) extracted from the $p-p$ momentum correlation function (Fig.~\ref{fig_Cpp_100MeV}), and the two-proton emission branching ratios ($B.R._{2p}$) (described in Section III). These results were obtained for 100 MeV $\gamma$ + $^{12}$C reactions. The experimental data for the RMS radius and $E_{bind}/A$ of $^{12}$C ground state are also listed.}
\label{f0table}
\begin{tabular}{ccccc}
\hline
      Configuration   ~~  & $r_{RMS}$   & ~~$E_{bind}/A$     & ~~$R_{source}$  & ~~$B.R._{2p}$     \\
   ~~  & (fm)  & ~~ (MeV)    & ~~ (fm) & ~~       \\ \hline
Chain &  2.71           & 7.17             & 1.85   & 0.45$\%$\\
Triangle & 2.35 	          & 7.12       &   1.55 & 0.75$\%$\\
Sphere    & 2.23          & 7.60              & 1.25 & 5.05 $\%$\\
Exp. Data &  2.4702(22)	           & 7.68       &  & \\  \hline
\label{table}
 \end{tabular}
\end{table}

\begin{table}[]
\centering
\caption{ Same as Table I but for  $^{16}$O configurations. }
\label{f0table}
\begin{tabular}{ccccc}
\hline
     Configuration   ~~  & $r_{RMS}$   & ~~$E_{bind}/A$     & ~~$R_{source}$  & ~~$B.R._{2p}$    \\
   ~~  & (fm)  & ~~ (MeV)    & ~~ (fm) & ~~       \\ \hline
Chain &  3.782           & 7.26      &    2.40  & 0.40$\%$\\
Kite  & 3.254 	          & 7.22      &   1.75  & 0.70$\%$\\
Square   & 2.908           & 7.29      &  1.60   & 0.85$\%$\\
Tetrahedron   & 2.761           & 7.79   &  1.50  & 1.30$\%$  \\
Sphere    & 2.6          &  8.15        &   1.40  & 5.13$\%$ \\
Exp. Data &  2.6991(52)	           & 7.976        \\  \hline
\label{table}
 \end{tabular}
\end{table}

From the given results of $C_{pp}$ as shown in Fig.~\ref{fig_Cpp_100MeV} for the 100-MeV photon energy case, emission source sizes of proton-proton pair ($R_{source}$) could be extracted. These are listed in Tables I and II. Traditionally, the source sizes are extracted by assuming the Gaussian source from the HBT correlation results. To accomplish this, the difference in emission times between two emitted protons should be considered as this is critical to obtain the correct source size. The Gaussian emission source in space and time can be written according to a function of $exp(-\frac{r^{2}}{2r^{2}_{0}} - \frac{t}{t_{0}})$, where $t_{0}$ is the lifetime for the emission of the second proton based on the assumption that the first proton is emitted at time $t$ = 0. We then obtain $t_{0}$ by fitting between $t$ and $t^{'}$, where $t$ is the distribution of emission times of the second proton, including all the events, and $t^{'}$ is sampled from a function of $exp(-\frac{t^{'}}{t_{0}})$. During the fit procedure, the best fitted radius of the source is obtained by searching a minimum of $\chi^{2}$ to fit the EQMD HBT results.

\begin{figure*}
\center
\includegraphics[scale=0.45]{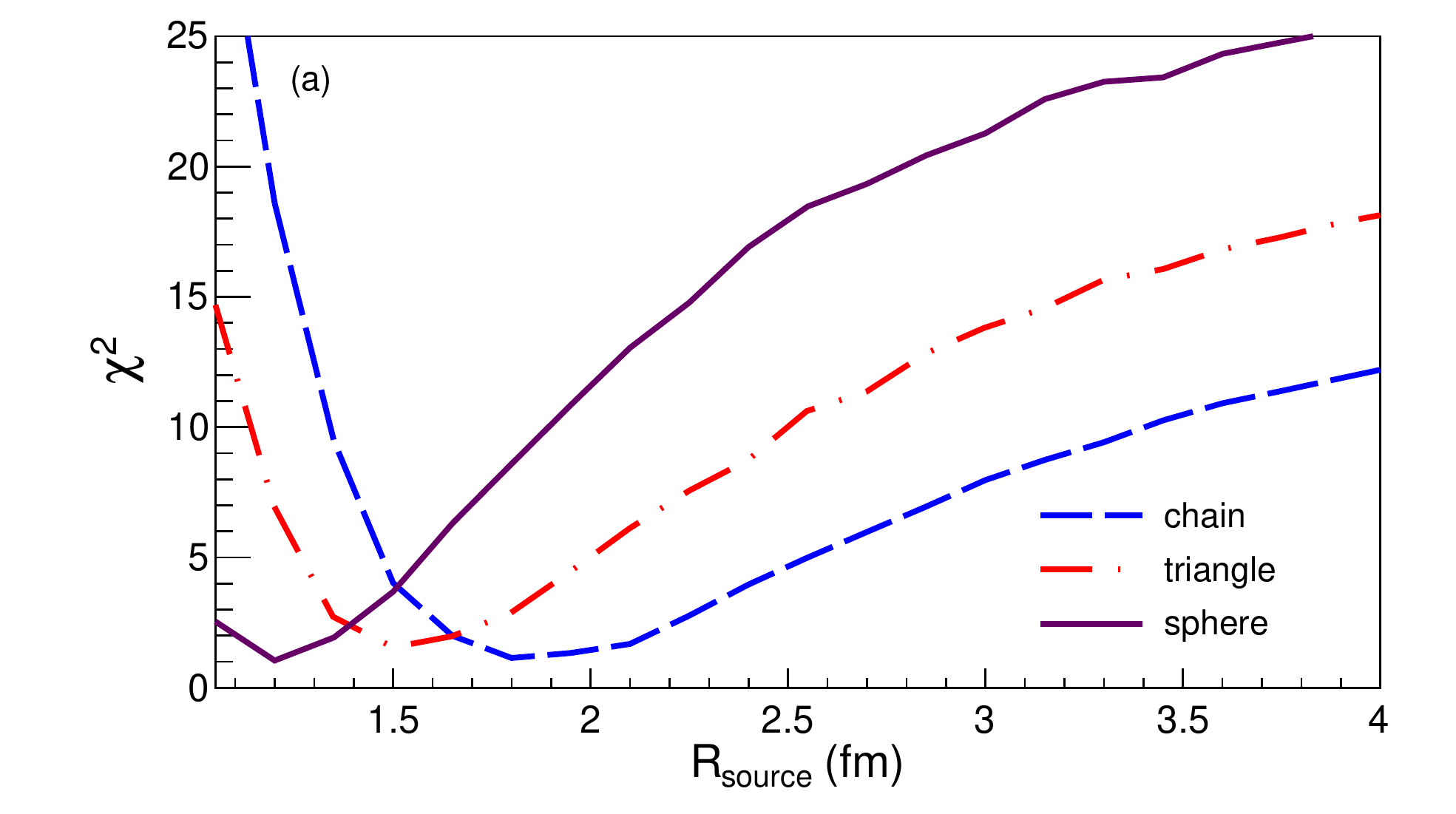}\includegraphics[scale=0.45]{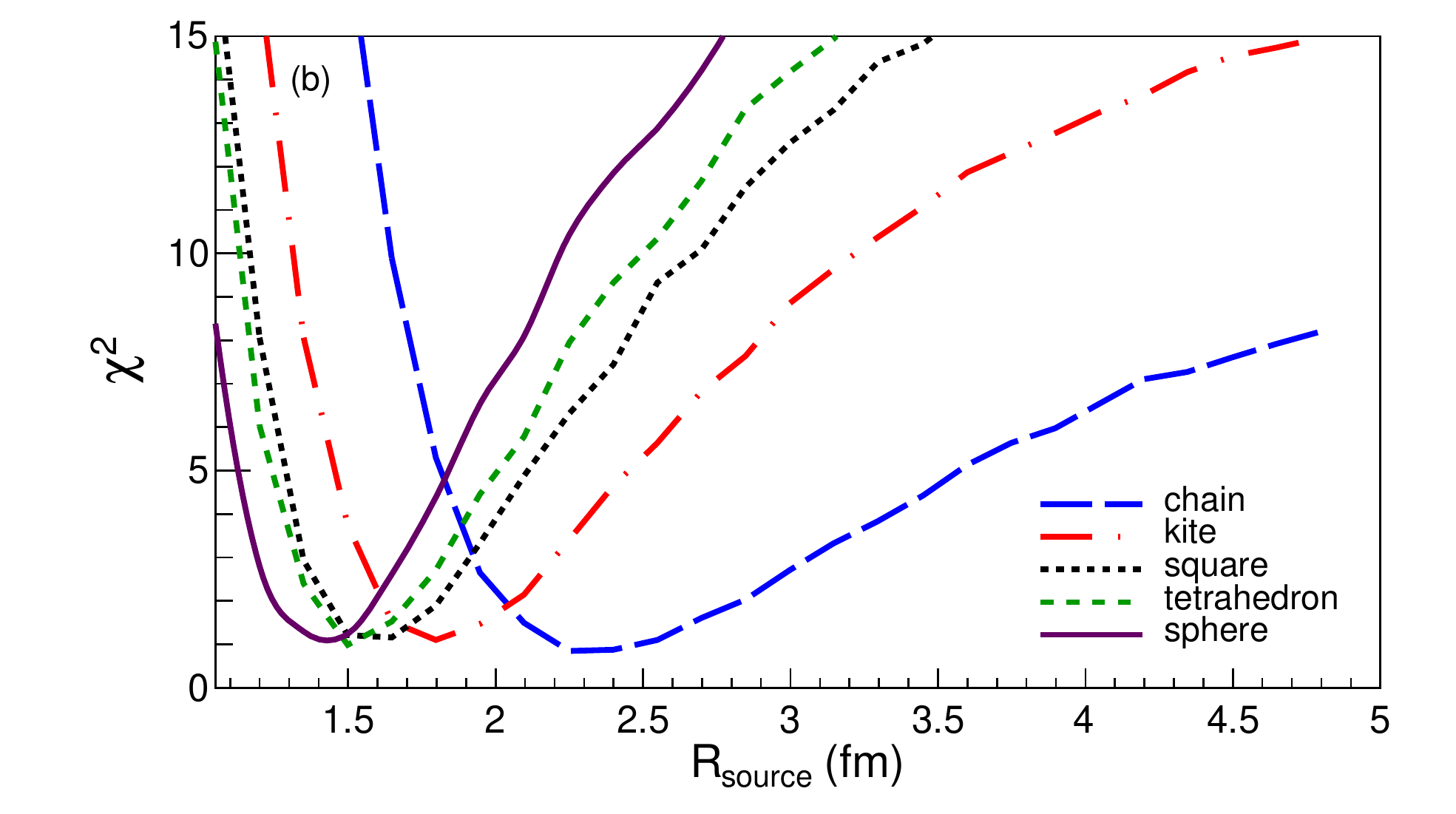}
\caption{ $\chi^2$ of the Gaussian source fits to proton-proton the momentum correlation functions shown in Fig. 1: (a) $^{12}$C, (b) $^{16}$O.}
\label{fig_size}
\end{figure*}

 Fig.~\ref{fig_size} shows the $\chi^{2}$ fits for the $p-p$ correlation function as a function of the radius of the Gaussian source for different configured structures
 of $^{12}$C (a) and $^{16}$O (b).
The points of minimum $\chi^{2}$ demonstrate that the chain configuration has the largest source size from among the different $\alpha$-clustering structures, the triangle and tetrahedron configurations have the minimum source sizes. Further, the kite and square configurations are between the chain and tetrahedron configurations for the $^{16}$O system. In addition, for the spheric nucleon distribution structure, the source size is the most compact.
 It is reasonable that the larger the space occupancy, the greater the size of the emission source from the proton-proton correlation functions. This indicates that the HBT technique is quite useful for reflecting the time-spatial structure, even for the exotic-shaped $\alpha$-clustering nuclei. These source sizes are listed in the fourth column in Tables I and II.

The fifth column in each of the two tables shows the branching ratios ($B.R._{2p}$) for the two-proton emission channel described in Section III. The tendency of $B.R._{2p}$ suggests that they are closely related to different configurations. The longer the RMS radius of the initial nucleus or the larger the proton-proton emission source size, then the lower the two-proton emission branching ratio. This phenomenon might be understandable based on a collision rate in space.

\begin{figure*}
\center
\includegraphics[scale=0.45]{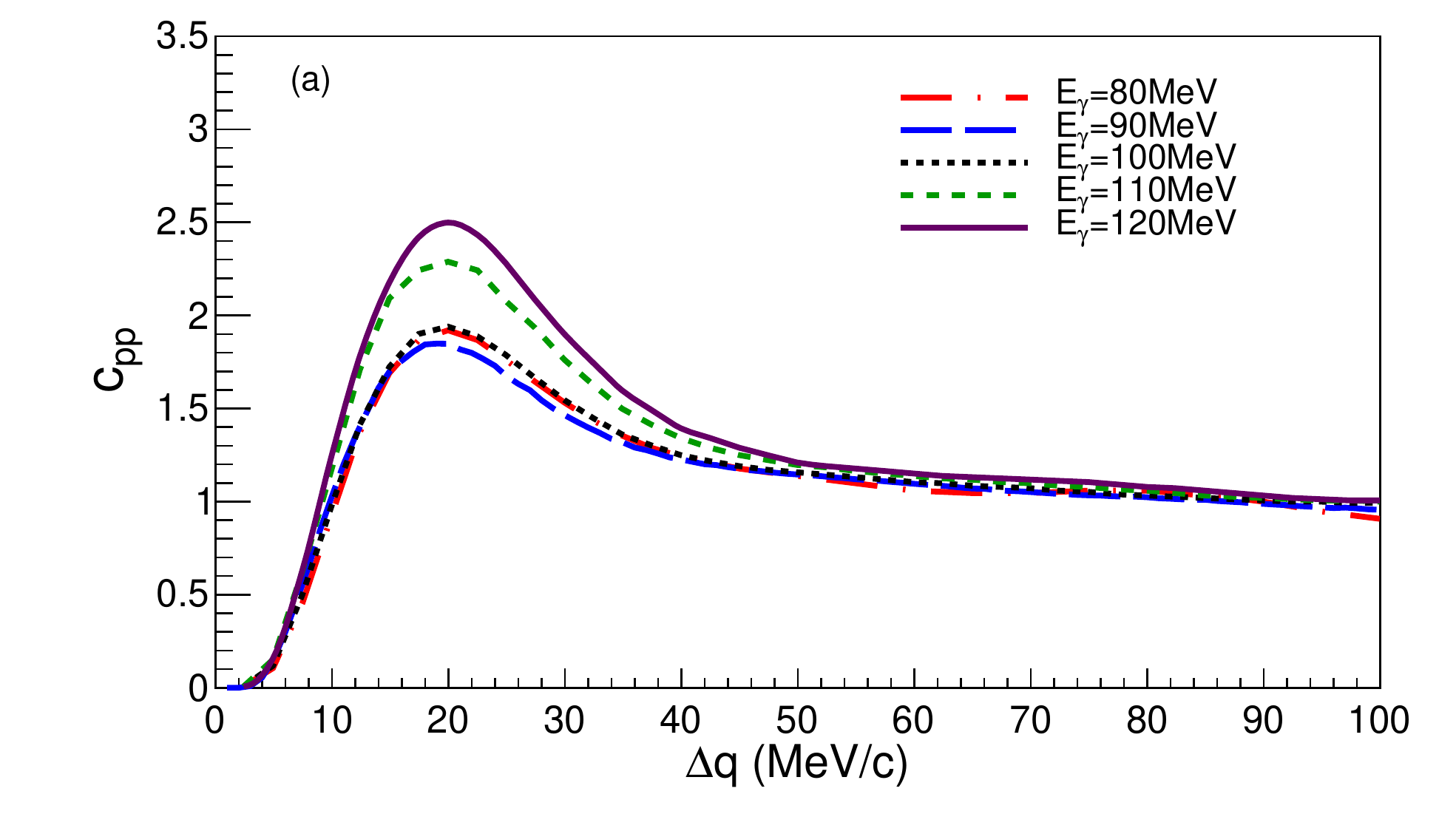}\includegraphics[scale=0.45]{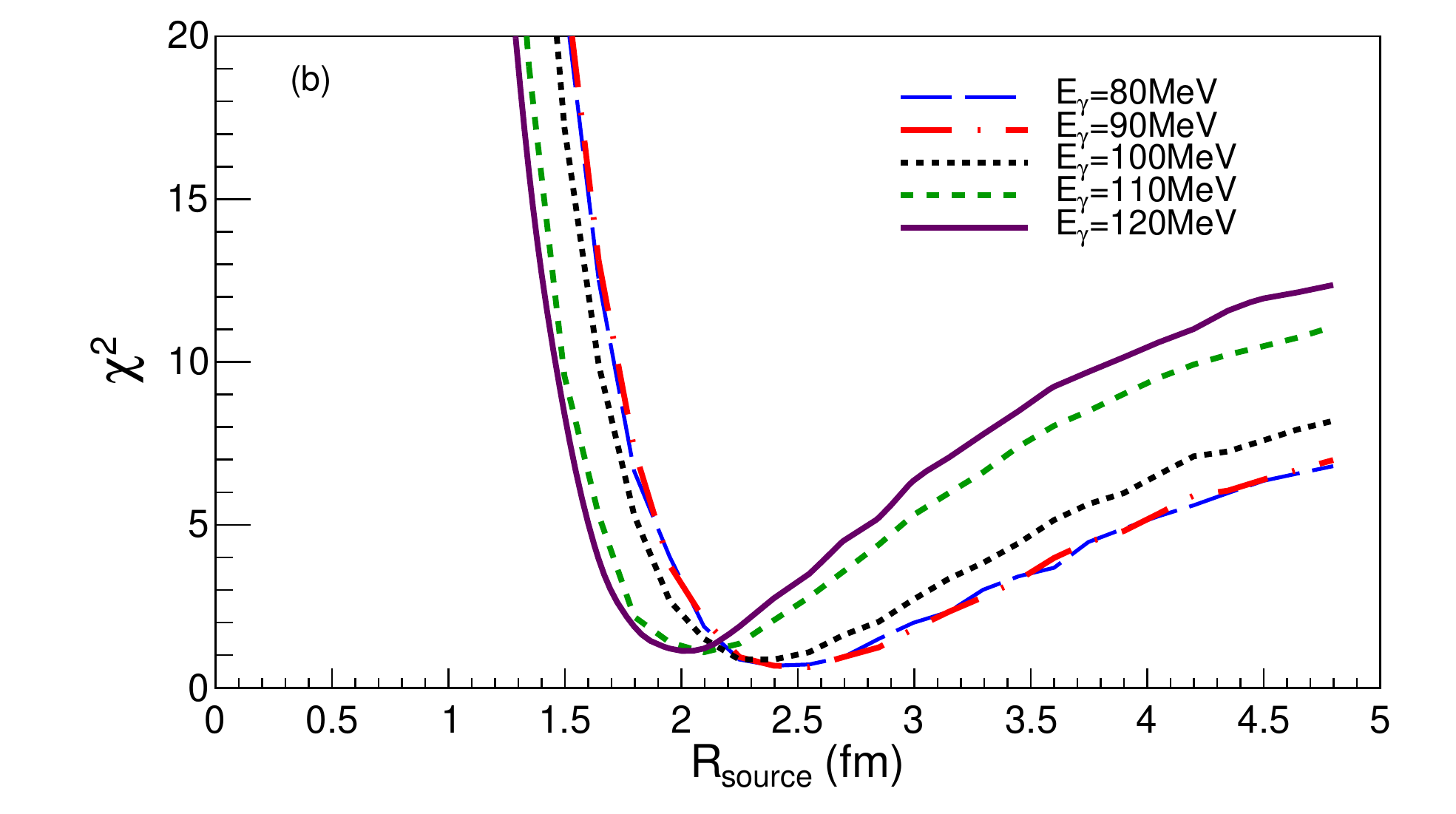}
\caption{Energy dependences of two-proton correlation functions (a) and the extracted source size (b) for $^{16}$O with chain four-$\alpha$ clustering structures.}
\label{fig_12C_Ebeam}
\end{figure*}

\subsection{Energy dependences of momentum correlation functions}

In previous studies, the photon energy was fixed at 100 MeV. Fig.~\ref{fig_12C_Ebeam} presents the correlation functions at different incident photon energies.
As an example, we show only the $p-p$ correlation functions for $^{16}$O with the chain four-$\alpha$ structure. In general, the figure displays sensitivities of the HBT strengths to photon energies, (i.e., stronger correlation at approximately 20 MeV/c emerges for higher incident energy). This may be explained by faster emission times for protons and/or more compact emission source sizes at higher photon energies. As a quantitative illustration, the right panel depicts the extracted source sizes at corresponding energies, revealing that the source sizes generally increase at lower incident energies. This is consistent with the HBT systematics with energy.

\section{Summary}

Three-body photo-disintegration channels from $^{12}$C($\gamma$,pp)$^{10}$Be and $^{16}$O($\gamma$,pp)$^{14}$C were investigated in a QD energy region within a framework of EQMD, and proton-proton momentum correlation functions were constructed and analyzed. In this study, phase-space information of nucleons at their emission times and the emission duration between two protons were extracted. Proton-proton momentum correlation functions were then obtained using the LL method for 100-MeV $\gamma$ + $^{12}$C and $^{16}$O targets, which were initialized by different geometric structures (i.e., random Woods-Saxon nucleon distribution and different $\alpha$-clustering structures).
For each nuclear configuration, the strength of the proton-proton momentum correlation function $C_{pp}$ demonstrated a sensitivity to the $\alpha$-clustering structure of $^{12}$C and $^{16}$O. This was also supported by the source sizes fitted by the Gaussian source to the momentum correlation functions.
The present work also determined that, in the QD regime, $C_{pp}$ is sensitive to incident photon energy, indicating that the emission source size depends on the photon energy. 

This study demonstrated that the construction of a proton-proton momentum correlation function is feasible in photo-nuclear reactions, and can be a promising tool for exploring nuclear structure information experimentally. In fact, high brilliance photon facilities such as HiGS \cite{Duke} and LEI-NP \cite{ELI} make this possible. In addition, these considerations could be applied to the (e,e$'$pp) reactions due to the availability of high-quality proton and electron beams \cite{Hen}. For future experimental studies investigating the $\alpha$-clustering structures of light nuclei, our study can shed light on the performance and momentum correlation analysis of ($\gamma$,pp) or (e,$e'$pp) reactions. 

\begin{acknowledgments}
This work is partially supported by the National Natural
Science Foundation of China under Contracts Nos. 11905284, 11890714,
and 11421505, the Key Research Program of Frontier
Sciences of the CAS under Grant No. QYZDJ-SSW-SLH002,
and the Strategic Priority Research Program of the CAS under
Grants Nos. XDPB09 and XDB16.

\end{acknowledgments}


\end{CJK*}


\begin{thebibliography}{99}


\bibitem{Ikeda}K. Ikeda, N. Takigawa, and H. Horiuchi, Prog. Theor. Phys. Suppl. E {\bf 68}, 464 (1968).

\bibitem{Greiner}W. Greiner, J. Y. Park, and W. Scheid, {\it Nuclear~ Molecules}  (World Scientific, Singapore, 1995).

\bibitem{Ortzen}W. von Oertzen, M. Freer, and Y. Kanada-Enyo, Phys. Rep. {\bf 432}, 43 (2006).

\bibitem{THSR}A. Tohsaki, H. Horiuchi, P. Schuck, and G. R\"opke, Phys. Rev. Lett.  {\bf 87}, 192501 (2001).

\bibitem{An}Zhen-Dong An, Yu-Gang Ma, Gong-Tao Fan, Yong-Jiang Li, Zhen-Peng Chen, and Ye-Ying Sun, The Astrophysical Journal Letters  {\bf 817}, L5 (2016).

\bibitem{An2}Zhen-Dong An, Zhen-Peng Chen, Yu-Gang Ma {\it et al.},  Phys. Rev. C  {\bf 92}, 045802 (2015).

\bibitem{Freer}M. Freer, Rep. Prog. Phys. {\bf 70}, 2149 (2007).

\bibitem{JBN}J. B. Natowitz, G. R\"opke, S. Typel  {\it et al.}, Phys. Rev. Lett. {\bf 104}, 202501 (2010).

\bibitem{Nature}J.-P. Ebran, E. Khan, T. Niksic, and D. Vretenar, Nature (London) {\bf 487}, 341 (2012).

\bibitem{Hoyle}F. Hoyle,   The Astrophysical Journal Supplement Series {\bf 1},  121 (1954).

\bibitem{Ichikawa}T. Ichikawa, J. A. Maruhn, N. Itagaki {\it et al.},  Phys. Rev. Lett.  {\bf 107}, 112501 (2011).

\bibitem{Suhara}T. Suhara, Y. Funaki, B. Zhou {\it et al.},  Phys. Rev. Lett.  {\bf 112}, 062501 (2014).

\bibitem{Schuck}M. Girod and P. Schuck, Phys. Rev. Lett. {\bf 111}, 132503 (2013).

\bibitem{16O_chain}T. Ichikawa, J. A. Maruhn, N. Itagaki, and S. Ohkubo, Phys. Rev. Lett.  {\bf 107}, 112501 (2011).

\bibitem{CFT1}E. Epelbaum, H. Krebs, D. Lee, and Ulf-G. Mei{\ss}ner, Phys. Rev. Lett.  {\bf 106}, 192501 (2011).

\bibitem{CFT2}E. Epelbaum, H. Krebs, T. A. L\"ahde, D. Lee, Ulf-G. Mei{\ss}ner, and G. Rupak, Phys. Rev. Lett.  {\bf 112}, 102501 (2014).

\bibitem{D3}D. J. Marn-Lmbarri, R. Bijker, M. Freer, M. Gai, Tz. Kokalova, D. J. Parker, and C. Wheldon, Phys. Rev. Lett.  {\bf 113}, 012502 (2014).

\bibitem{Zhou}Bo Zhou, Y. Funaki, H. Horiuchi, Zhongzhou Ren, G. R\"opke, P. Schuck, A. Tohsaki, Chang Xu, and T. Yamada, Phys. Rev. Lett. {\bf 110}, 262501 (2013).

\bibitem{CaoXG-2019}X. G. Cao, E. J. Kim, K. Schmidt {\it et al.}, Phys. Rev. C {\bf 99}, 014606 (2019).

\bibitem{NST}Y. Liu and Y. L. Ye, Nucl. Sci. Tech. {\bf 29}, 184 (2018).

\bibitem{Enyo} Y. Kanada-Enyo,  M. Kimura, F. Kobayashi, T. Suhara, Y. Taniguchi, and Y. Yoshida, Nucl. Sci. Tech. {\bf 26}, S20501 (2015).

\bibitem{W.B.He} W. B. He, Y. G. Ma, X. G. Cao, X. Z. Cai and G. Q. Zhang, Phys. Rev. Lett. {\bf 113}, 032506 (2014); ibid, Phys. Rev. C {\bf 94}, 014301 (2016).

\bibitem{Guo}Chen-Chen Guo, Yu-Gang Ma, Zhen-Dong An,  and Bo-Song Huang, Phys. Rev. C {\bf 99}, 044607 (2019).

\bibitem{ZhangS}S. Zhang, Y. G. Ma, J. H. Chen, W. B. He, C. Zhong, Phys. Rev. C {\bf 95}, 064904 (2017); ibid, Eur. Phys. J. A {\bf 54}, 161 (2018).

\bibitem{XuZW}Zhi-Wan Xu, Song Zhang, Yu-Gang Ma, Jin-Hui Chen, Chen Zhong, Nucl. Sci. Tech. {\bf  29}, 186 (2018).

\bibitem{Nuclear photonfissility} O. A. P. Tavares, S. B. Duarte, A. Deppman $et \,al.$, J. Phys. G {\bf 30}, 377 (2004).

\bibitem{Duke}H. R. Weller, A. W. Mohammad, H.  Gao  {\it et al.}, Prog. Part. Nucl. Phys.   {\bf  62},    257 (2009).

\bibitem{SIOM}C.  Yu, R. Qi,  W.  Wang  {\it et al.}, Sci. Rep. {\bf 6}, 29518  (2016).

\bibitem{ELI}D. Filipescu, A. Anzalone, D. L. Balabanski  {\it et al.}, Eur. Phys. J. A {\bf 51}, 185   (2015).

\bibitem{SINAP}H. L. Wu, J. H. Chen, B. Liu  {\it et al.}, Nucl. Sci. Tech. {\bf  26},    050103 (2015).

\bibitem{Amano} S. Amano, K. Horikawa, K. Ishihara  {\it et al.},  Nucl. Inst. Meth. A {\bf 602} , 337 (2009).

\bibitem{photo_16O}A. Leistenschneider {\it et al.}, Phys. Rev. Lett. {\bf 86}, 5442 (2001).

\bibitem{J.S.Levinger0} J. S. Levinger, Phys. Rev. {\bf 84}, 43 (1951).

\bibitem{np} E. C. Sympson and J. A. Tostevin, Phys. Rev. C {\bf 83}, 014605 (2011).


\bibitem{huang12C} B. S. Huang, Y. G. Ma, W. B. He, Phys. Rev. C {\bf 95}, 034606 (2017).

\bibitem{huang16O} B. S. Huang, Y. G. Ma, W. B. He, Eur. Phys. A {\bf 53}, 119 (2017).


\bibitem{Hanbury} R. Hanbury Brown, R. Q. Twiss, Nature {\bf 178}, 1046 (1956).

\bibitem{Goldhaber} G. Goldhaber $et \,al.$, Phys. Rev. {\bf 120}, 300 (1960).

\bibitem{myg} Y. G. Ma,Y. B. Wei, W. Q. Shen $et \,al.$, Phys. Rev. C {\bf 73}, 014604 (2006).

\bibitem{cxg} X. G. Cao, Y. G. Ma, D. Q. Fang $et \,al.$, Phys. Rev. C {\bf 86}, 044620 (2012).

\bibitem{wtt} T. T. Wang, Y. G. Ma, C. J. Zhang, Z. Q. Zhang, Phys. Rev. C {\bf 97}, 034617 (2018).

\bibitem{wtt2} T. T. Wang, Y. G. Ma, Z. Q. Zhang, Phys. Rev. C{\bf  99}, 054626 (2019).

\bibitem{STAR} L. Adamczyk $et\, al.$ (STAR Collaboration), Nature {\bf 527}, 345 (2015).

\bibitem{ChenJH}Jinhui Chen, Declan Keane, Yu-Gang Ma, Aihong Tang, Zhangbu Xu, Phys. Rep. {\bf  760}, 1 (2018).

\bibitem{Neha}J. Adam {\it et al.} (STAR Collaboration), Phys. Lett. B {\bf 790}, 490 (2019).

\bibitem{WangFan}T. Goldman, K. Maltman, G.J. Stephenson, K.E. Schmidt, F. Wang, Phys. Rev. Lett. {\bf 59}, 627 (1987).

\bibitem{ShenPN}Q. B. Li, P. N. Shen, Z. Y. Zhang, Y. W. Yu, Nucl. Phys. A {\bf 683}, 487 (2001).

\bibitem{Xi} Bao-Shan Xi, Zheng-Qiao Zhang, Song Zhang, and Yu-Gang Ma, arXiv:1909.03157v1.

\bibitem{Ma2015} Y. G. Ma, D. Q. Fang, X. Y. Sun {\it et al.}, Phys. Lett. B {\bf 743}, 306 (2015).

\bibitem{Fang2016} D. Q. Fang, Y. G. Ma,  X. Y. Sun {\it et al.}, Phys.  Rev. C {\bf 94}, 044621 (2016).

\bibitem{Wang2018} Yu-Ting Wang,  De-Qing Fang, Xin-Xing Xu, Kang Wang {\it et al.}, Phys. Lett. B {\bf 784}, 12 (2018).

\bibitem{Wang2018B} Yu-Ting Wang,  De-Qing Fang, Xin-Xing Xu {\it et al.}, Nucl. Sci. Tech. {\bf 29}, 98 (2018).

\bibitem{Marques} F. M. Marques, M. Labiche, N.A. Orr {\it et al.}, Phys. Rev. C {\bf 64},
061301 (2001).

\bibitem{Wei}Y. B. Wei, Y. G. Ma, W. Q. Shen {\it  et al.}, Phys. Lett. B {\bf 586}, 225 (2004).

\bibitem{MARUYAMA} T. Marayama, K. Niita, A. Iwamoto,  {Phys. Rev. C} {\bf 53},  297 (1996).

\bibitem{J.Aichelin} J. Aichelin, H. Stocker, Phys. Lett. B {\bf 176}, 14 (1986).

\bibitem{J.Aichelin2} J. Aichelin,  Phys. Rep. {\bf 202}, 233 (1991).

\bibitem{C.Hartnack} C. Hartnack, R. K. Puri, J. Aichelin $et \,al.$, Eur. Phys. J. A {\bf 1}, 151 (1998).

\bibitem{C.Hartnack1} C. Hartnack, Zhuxia Li, {Nucl. Phys. A} {\bf 495}, 303 (1989).

\bibitem{FengNST}
Z. Q. Feng, Nucl. Sci. Tech. {\bf 29}, 40 (2018).

\bibitem{A.Ohnishi} A. Ohnishi, T. Maruyama, H. Horiuchi, {Prog. Theor. Phys} {\bf 87}, 417 (1992).


\bibitem{P.Valta} P. Valta, J. Konopka, A. Bohnet $et al.$, {Nucl. Phys. A} {\bf
538}, 417 (1992).


\bibitem{A.Ono} A. Ono, H. Horiuch, T. Maruyama $et\ al.$, {Prog. Theor. Phys} {\bf
87}, 1185 (1992).


\bibitem{J.S.Levinger}J. S. Levinger,  {Phys. Lett. B} {\bf 82}, 181 (1979).


\bibitem{P.Rossi}P. Rossi, E. De Sanctis, P. Levi Sandri {\it et al.},  {Phys. Rev. C} {\bf
40}, 2412 (1989).


\bibitem{Lednicky}R. Lednicky, Sov. J. Nucl. Phys. {\bf 35}, 770 (1982).

\bibitem{Koonin1977}S. E. Koonin, Phys. Lett. B {\bf 70}, 43 (1977).

\bibitem{Lednicky1} R. Lednicky, V. L. Lyuboshitz, B. Eranmus {\it et al.},  Phys Lett B {\bf 373}, 30 (1996).

\bibitem{Lednicky2009} R. Lednick$\acute{y}$, Phys. Part. Nucl. {\bf 40}, 307 (2009).

\bibitem{Lednicky2008} R. Lednick$\acute{y}$, Phys. Ato. Nucl. {\bf 71}, 1572 (2008).

\bibitem{Erazmus1994} B. Erazmus, L. Martin, R. Lednicky, N. Carjan, Phys. Rev. C {\bf 49}, 349 (1994).

\bibitem{Arvieux1974} J. Arvieux, Nucl. Phys. A {\bf 221}, 253 (1974).

\bibitem{McGeorge}
J. C. McGeorge, I. J. D. MacGregor, S. N. Dancer {\it et al.}, Phys. Rev. C {\bf 51}, 1967 (1995).

\bibitem{McGeorge2}I. J. D. MacGregor {\it et al.}, Nucl. Phys. A {\bf 533}, 269 (1991).

\bibitem{MacNew}
I. J. D. MacGregor,
SciPost Physics Proceedings; 24th European Few Body Conference (University of Surrey, U.K.).

\bibitem{Hen}
Or Hen, Gerald A. Miller, Eli Piasetzky, Lawrence B. Weinstein, 
Rev. Mod. Phys. {\bf 89}, 045002 (2017).


\end{thebibliography}
\end{document}